# A New Supersensitive Flame Detector and it's Use for Early Forest Fire Detection

V. Peskov and A. Zichichi

LAA-Project-CERN, Geneva-Switzerland
E. Fermi-Center, Rome-Italy

## Abstract

A new flame detector, three orders of magnitude more powerful than the existing ones, is presented. This detector needs to be mass-produced for its use in order to be incorporated in an early forest fire detection system. A project able to implement it's use to overcome the forest fire emergency is described.

# I. Introduction

Forests fires, which were especially strong this summer, are not only dangerous for human life, people properties and industrial infrastructure, but also may strongly damage our environment. Indeed forests are the most important air cleaning factory: they act as carbon sinks and as physical air filter. Data on forest fire and consequent burned area in Mediterranean countries and its neighbors in South Europe are shown in Table 1(from J. Goldammer et al., presented at the International Conference "Fire and emergency safety", October 2002 , Bulgaria). One can clearly see that the loses affect many countries including Italy.

Table 1. Fire statistical data of the Mediterranean countries and its neighbors in South Europe

| Country | Time period | Average number of fires | Total area burned, ha |
|---|---|---|---|
| Albania | 1981-2000 | 667 | 21456 |
| Algeria | 1979-1997 | 812 | 37037 |
| Bulgaria | 1978-1990 | 95 | 572 |
|  | 1991-2000 | 318 | 11242 |
| Cyprus | 1991-1999 | 20 | 777 |
| Croatia | 1990-1997 | 259 | 10000 |
| France | 1991-2000 | 5589 | 17832 |
| Greece | 1990-2000 | 4502 | 55988 |
| Israel | 1990-1997 | 959 | 5984 |
| **Italy** | **1990-1999** | **111163** | **118576** |
| Lebanon | 1996-1999 | 147 | 2129 |
| Morocco | 1960-1999 | n.a. | 2856 |
| Portugal | 1990-1997 | 20019 | 97175 |
| Romania | 1990-1997 | 102 | 355 |
| Slovenia | 1991-1996 | 89 | 643 |
| Spain | 1990-1999 | 18105 | 159935 |
| Turkey | 1990-1997 | 1973 | 11696 |

Thus early forest fire alarm system is an urgent need for Italy and other European countries. We propose to use advanced technology in order to provide an efficient early forest fire alarm system. The fundamental component of the system is a very efficient detector. In §II we describe its origin and its technological structure. Its basic property are in §III and in §IV we discuss what needs to be done in order to go from the existing powerful detector (very sensitive to small flames even in fully illuminates areas) to a system for an early forest fire alarm. The estimated costs for each detector and the budget for its mass-production are in §V and §VI respectively; the conclusions in §VII.

**II. What has been done in the frame of the LAA project**

In the frame of the LAA project [1-12] we have developed a novel photosensitive detector for BaF$_2$ calorimetry. It is schematically shown in Fig.1

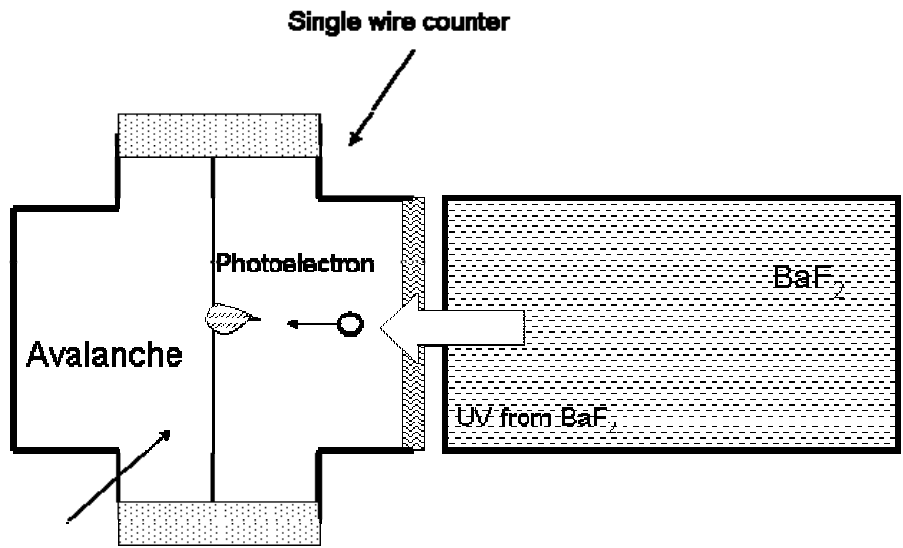

Fig. 1. The schematic drawing of the gaseous detector filled with photosensitive vapors: EF or TMAE (LAA project)

The principle of its operation is based on photoionzation of gases with small ionization potential. For BaF$_2$ calorimetry we successfully tested and used several different photosensitive vapors, for example TMAE and Ethylferrocene (EF). We also successfully tested various solid photocathodes [10] including CsI [11].

As an illustration in Fig. 2 a schematic drawing of one of our first original device developed in the frame of the LAA project (see the paper by V. Peskov and A. Zichichi, Preprint CERN-PPE/91-139) is shown [12].

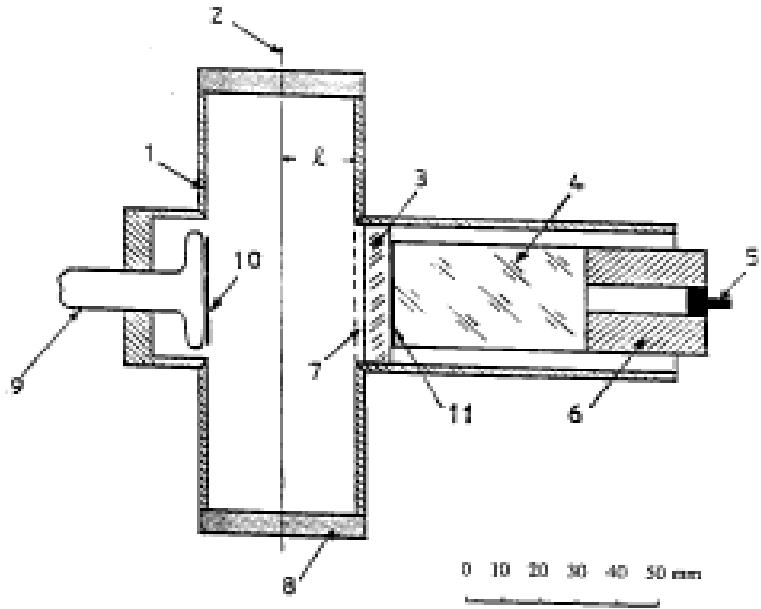

Fig. 2. A drawing of the UV sensitive detector developed in the frame of the LAA project. 1-body of the counter (grounded), 2- anode wire to which the high voltage is applied, 3-$CaF_2$ window, 4-crystal to be tested, 5-radioactive source, 6- lead collimator, 7- cathode mesh, 8-ceramics, 9-aluminium cathode, 10-replacable cathode, 11-optical contact

Note that the detector shown if Fig. 2 is just a single example of many UV detectors tested.

We have recently discovered that the same detector can be used for detection of small flames in fully illuminated areas. The reason why this device can detect even a small flame is the following. The sunlight in the wavelength interval 185-280 nm is fully absorbed in the higher part of the atmosphere by the ozone layer, while the atmosphere is transparent for these wavelengths at the ground level. On the other hand, all flames in air emit in the wavelength interval 185-260 nm. This offers a unique possibility to detect flames and fires without background from the visible and UV light from the Sun and from the visible light in a room. Our extensive tests show that any of photosensitive elements developed in the frame of the LAA project can be used for this particular application (TMAE, EF, CsI and so on), however the best is TMAE vapors: they have the highest sensitivity to flames and at the same time are not sensitive to the direct sun light (see Fig. 3).

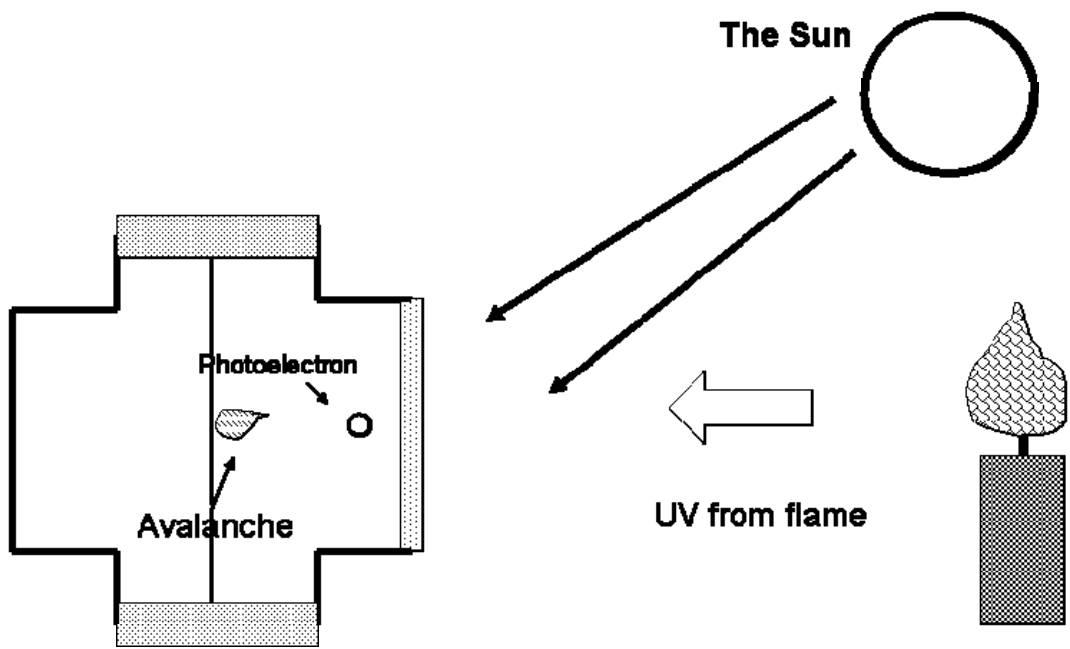

Fig. 3. A schematic drawing of the LAA detector filled with TMAE photosensitive vapors able to detect small flames. Our detector is not sensitive to the direct sun light and also to the light in a room

We have developed several prototypes of flame detector (see for example Fig. 4)

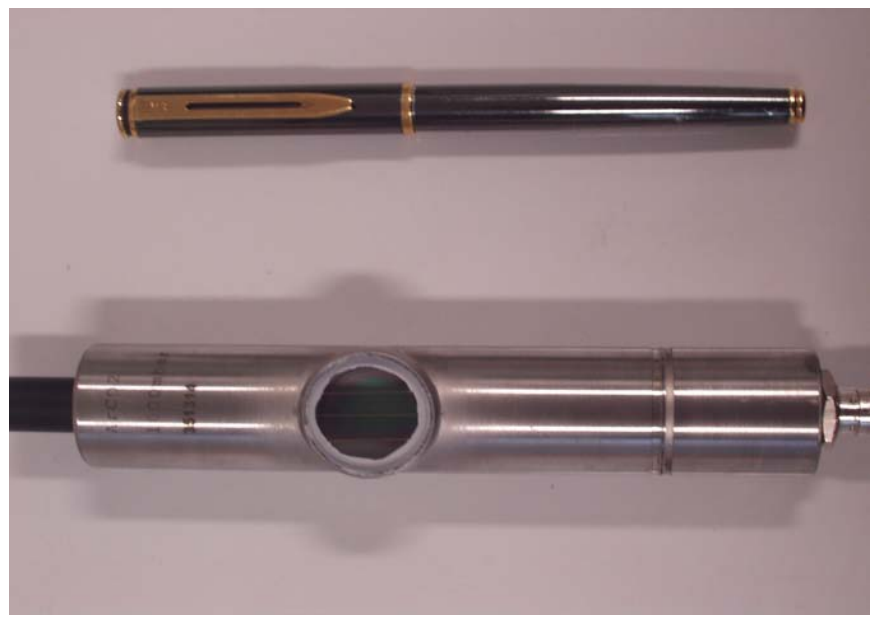

Fig. 4. A photo of the sealed flame detector

and our extensive tests fully confirm that this detector is able to detect very small flames in open air in the presence of the direct sunlight and in a fully illuminated room. Therefore we would like to propose to organize a mass production of this original device which has remarkable advantages for the forest fire detection as well as for detection of fires in buildings.

**III. Basic characteristics of our flame detector**

1. Sensitivity

The sensitivity is of course the most critical issue.
One should note that there are commercially available UV detectors of flame. They are classified by three EU standards (EN54-10 2002 (E)); the highest is called "class 1". The requirement for the class 1 detectors is a capability to detect a flame with a size of 30x30x30cm$^3$ during 30 sec after its appearance at a distance of 25m from the sensor. Our detectors operating on a different principle than the commercial ones are over 1000 time more sensitive. For example a small flame from a cigarette lighter can be easily detected at a distance of 30m
To confirm these qualitative results we have performed direct quantative comparison with the 1$^{st}$ class Hammatsu flame detector R2868. The results are summarized in <u>Table 2.</u> From the data presented (number of counts per 10sec) it is clearly seen that <u>our detector is over 1000 time more sensitive than the first class commercial detectors.</u>

| Hamamatsu R2868 | | Our detector | |
| --- | --- | --- | --- |
| Distance (m) | Mean number of counts per 10sec: $N_H$ | Mean number of counts per 10sec: $N_O$ | Ratio $N_O/N_H$ |
| 1,1 | 583 | 690747 | 1.18 x 10$^3$ |
| 3 | 76 | 91013 | 1.19 x 10$^3$ |
| 10 | 6 | 7820 | 1.30 x 10$^3$ |
| 30 | 0.1 | 873 | 8 x 10$^3$ |
| 85 | ~0.1 | 51 | |

Table 2. Comparison of number of counts detected from the Hamamatsu R2868 flame sensor and our detector, both detecting a flame from the candle in fully illuminated areas. Note that at distances ≥30 m the counting rate from the Hamamatsu device was just due to its internal noise

Note that the counting rate without candle was 20 per 10sec and was mainly caused by the cosmic radiation.

2. Time characteristics

The other unique feature of our detectors is that, in contrast to commercial devices, our detectors react to a flame in a few μs (so it is $10^7$ times faster than required by the EU standards (EN54-10 2002 (E)) and this allows to expand the possible fields of application, for example to spark detection in dangerous places like oil drill platform, lighting detection and many others.

3. Power consumption

The third important characteristic of the flame detector is the power consumption.
Our detector itself does not consume practically any power ($<10^{-3}$W); the only power is consumed by the electronics. Modern electronics can be easily feed by a compact battery, so the power should be: 10mAx12V=0.12W

4. Dimensions.

At present the sensitive element has the following dimensions: diameter 25 mm, length 170mm. We estimate that with a compact electronic and HV supplier the total dimensions of the detector will be less than 70x70 x190mm$^3$.

5. Temperature range

Our detector can operate in the temperature interval from minus ten to hundred degrees Centigrade. If necessary this temperature interval can be expanded. The present prototypes can be destroyed in fire, however if necessary one can design a detector withstanding few hundreds degrees Centigrade.

6. Humidity

The present designs work well at humidity <80%, but this could be easily expanded to 100%.
We are planning to make future detector designs fully waterproof allowing them to operate during rains.

**IV. What is needed**

IV.1. Product industrialization
In order to have a system for forest early fire detection it is necessary to study the product industrialization. For this purpose it is necessary:
a) To develop a compact electronics for the detector.
b) To make a final industrial prototype equipped with compact electronics.
c) To test new detectors in real environment: for detection of flames occurring on a long distance and at any weather.

d) To develop a system comprising our device and pulsed compact UV sources for monitoring the smoke and dangerous vapors which may appear before or after the flame as illustrated in Fig. 5.
These pulsed UV sources will greatly expand the capability of our detector.

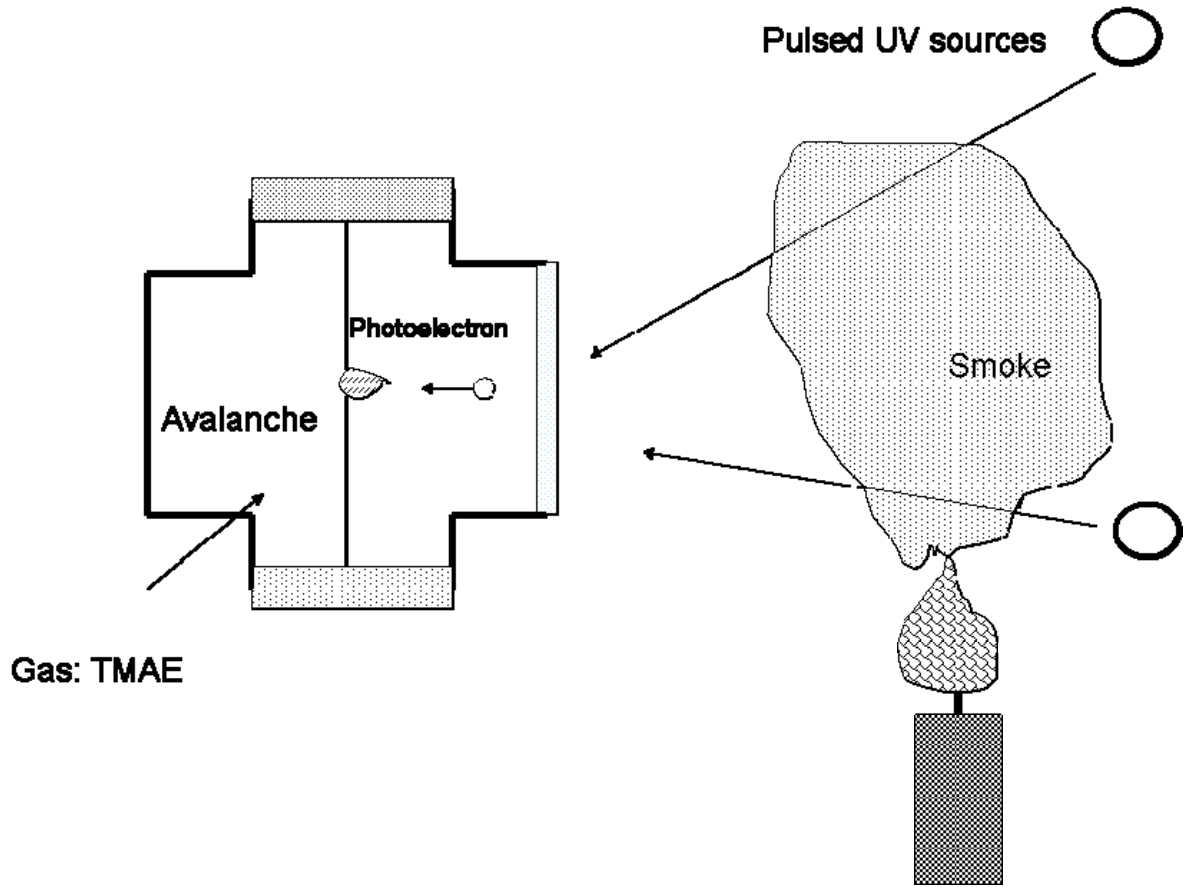

Fig. 5. A schematic view of our detector combined with external pulsed UV source, which allows continuous monitoring of the atmosphere transmission and thus detect appearance of any smoke.

IV.2. Mass production and applications

Priorities in this context are as follows:
a) The basic element of the detector (its sensitive element) should be pumped to a high vacuum and then filled with photosensitive vapors. A small production line should be build to start with.
b) The second step is a mass production of these new very high sensitive detectors oriented on their application for flame detection both for forest fire detection and for detection of small flames and fires inside buildings.
The mass production needs using the big laboratory infrastructures at CERN.
c) Once the region of Europe is chosen, a network of towers in this region should be equipped with our detectors, feed either by usual batteries or by solar panels and having a radio communication with the fire safety headquarter (see Fig. 6).

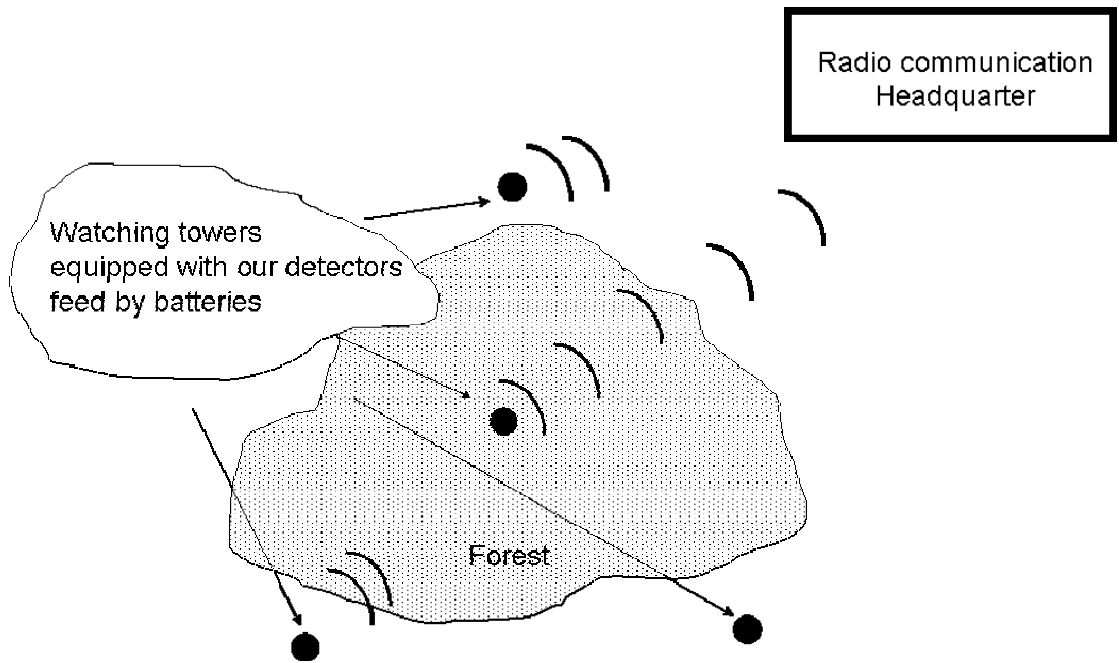

Fig .6. A schematic drawing of the network of watching towers with a radio communication system forest fire detection.

## V. Estimated dimensions and cost per detector

The estimated dimensions of our detector for indoor applications will be 70x70x70 mm$^3$, its cost will not exceed100 Euro. Note that the cost of the Hamamatsu R 2868 flame sensor is around 70 Euro (quotation from Hamamatsu obtained in 2005).
The estimated cost of the solar battery feed detectors for forest fire (equipped with a radio transmitter) is at most 200 Euro per detector.
We believe that after further development the cost can be reduce by at least a factor of two.

## VI. Estimated budget to implement the mass production project

1. Technical Developments

To develop an industrial prototype equipped with compact integrated electronics and a pulsed UV sources: 50 kEuro.
To set a small production line of flame detectors at CERN: 250 kEuro.
Materials: 100 kEuro.
Total: 400 kEuro.

2. Personnel

Physicist: 65 kEuro per/year x 2years= 130 kEuro
Electronics Engineer: 50 kEuro per/year x 2years=100 kEuro
Technician: 40 kEuro per/year x 2years= 80 kEuro
Total: 310 kEuro.

3. The cost of the tower network for the forest fire alarm system is to be added once the area to be operated is chosen.

Total for the entire project: 710 kEuro.

## VII. Conclusions

The innovative principle which is the basic feature of our new supersensitive flame detector has been clearly proven. We have experimentally established that our detector:
a) is 1000 times more sensitive than the best commercial devices;
b) can detect not only small flames and fires, but sparks as well;
c) can be battery feed;
d) can operate in direct sunlight.

These unique features motivated us to study this detector as basic instrument for early forest fire detection in order to solve the problem of forest fire. What needs to be done is the study of its mass production and of few other details related to this specific use.